\documentclass[
    ,final            
  ]
  {myaipproc}

\layoutstyle{6x9}

\begin{document}

\title{Dynamics study of $Z^+(4430)$ and $X(3872)$ in molecular picture}


\author{Xiang Liu \footnote{e-mail: xiangliu@pku.edu.cn, liuxiang@teor.fis.uc.pt}}{
  address={Department of Physics, Peking University, Beijing
100871, China \\ Centro de F\'{i}sica Te\'{o}rica, Departamento de
F\'{i}sica, Universidade de Coimbra, P-3004-516, Coimbra,
Portugal}}

\author{Yan-Rui Liu}{address={Institute of High Energy Physics, P.O. Box 918-4, Beijing
100049, China}}

\author{Wei-Zhen Deng}{address={Department of Physics, Peking University, Beijing
100871, China}}

\begin{abstract}
In this talk, we review our recent work about the dynamical
studies of $Z^+(4430)$ and $X(3872)$. $Z^+(4430)$ can not be
explained as a $D_1'D^*$ or $D_1D^*$ molecular state only
considering one pion exchange potential without the cutoff, which
needs to be confirmed by introducing sigma exchange potential and
adding the cutoff in the effective potential. One also excludes
the possibility of $X(3872)$ as a $DD^*$ molecular state by one
pion and one sigma exchanges with the cutoff. Fortunately there
exists an S-wave $BB^*$ bound state with $J^{PC}=1^{++}$. we
suggest future experiment to search this state.
\end{abstract}

\maketitle

\section{Introduction}

$Z^{+}(4430)$, a new enhancement announced by Belle Collaboration,
has stimulated theorists to speculate its underlying structure.
Its mass and width are respectively
$m=4433\pm4(\mathrm{stat})\pm1(\mathrm{syst})$ MeV and $
\Gamma=44^{+17}_{-13}(\mathrm{stat})^{+30}_{-11}(\mathrm{syst})$
MeV. The isospin and G-parity of $Z^+(4430)$ are $I^G=1^+$ because
$Z^+(4430)$ was observed in the $\psi'\pi^+$ channel
\cite{Belle-4430}. The explanations for its structure mainly
include the S-wave threshold effect \cite{rosner}, the $D_1D^*$
molecular state \cite{Meng,ding,qsr}, the tetraquark state
\cite{maiani,cky,Gershtein}, the cusp effect \cite{Bugg} and the
$\Lambda_c-\Sigma_c^0$ bound state \cite{Qiao}. Our recent work
\cite{xiangliu} reviewed the recent theoretical status of
$Z^+(4430)$
\cite{rosner,Meng,ding,qsr,maiani,cky,Gershtein,Bugg,Qiao,braaten}
and explored whether $Z^+(4430)$ can be explained as $D_1'D^*$ or
$D_1D^*$ molecular state by one pion exchange (OPE).

In a series of XYZ charmonium-like states observed in recent
years, $X(3872)$ \cite{3872} is also a state near the threshold of
$DD^*$, which attracted extensive concerns of theorists
\cite{3872-Mole-1,3872-Mole-2,3872-Mole-3,3872-Mole-4,3872-Mole-5,3872-cusp,3872-s-wave,3872-hybrid,3872-D,3872-tetra}.
Among these theoretical explanations, the molecule picture is the
most popular one
\cite{3872-Mole-1,3872-Mole-2,3872-Mole-3,3872-Mole-4,3872-Mole-5}
even though the predictions in $DD^*$ molecule picture are
inconsistent with the experimental measurements to some extent. In
fact, only dynamics studies can give a reasonable answer about
whether $X(3872)$ can be interpreted as a $DD^*$ molecular state.
Swanson proposed that $X(3872)$ was mainly a $D^0 \bar{D}^{*0}$
molecule bound by both the pion exchange and quark exchange
\cite{3872-Mole-4}. In Ref. \cite{3872-Mole-3}, Wong studied the
$DD^*$ system in the quark model in terms of a four-body
non-relativistic Hamiltonian with pairwise effective interactions,
and found an S-wave $DD^*$ molecule with the binding energy $\sim
7.53$ MeV. However, with the obtained one pion exchange potential
(OPEP) by using the effective Lagrangian, Suzuki argued that
$X(3872)$ is not a molecular state of $D^0\bar{D}^{*0}+\bar{D}^0
D^{*0}$ \cite{suzuki}, which contradicts Swanson and Wong's
conclusion. In our recent work, we reexamined whether $X(3872)$ is
a molecular state by adding $\sigma$ meson exchange potential and
introducing the cutoff in the effective potential \cite{3872-liu}.

In this talk, we will briefly introduce the theoretical framework
of deducing the effective potential. Then we will respectively
discuss whether $Z^+(4430)$ can be $D_1'D^*$ or $D_1D^*$ bound
state and whether $X(3872)$ can be explained as $DD^*$ molecular
state based on our recent work presented in Ref.
\cite{xiangliu,3872-liu}. In the last section, a summary will be
given.

\section{Theoretical Framework}

To derive the effective potential, firstly we write out the
elastic scattering amplitudes of system according to the
Lagrangian, which is constructed in the chiral and heavy quark
dual limits \cite{falk,casalbuoni}
\begin{eqnarray}
\mathcal{L}&=&ig {\rm Tr}[H_b
{A}\!\!\!\slash_{ba}\gamma_5\bar{H}_a ]+ig'{\rm Tr}[ S_b
{A}\!\!\!\slash_{ba}\gamma_5\bar{S}_a]
 +ig''{\rm Tr}[T_{\mu b}
A\!\!\!\slash_{ba}\gamma_5\bar{T}_a^{\mu}]\nonumber\\&&+[ih {\rm
Tr}[S_b{A}\!\!\!\slash_{ba}\gamma_5
\bar{H}_a]+h.c.]+\{i\frac{h_1}{\Lambda_{\chi}}{\rm
Tr}[T_b^{\mu}(D_{\mu}{A}\!\!\!\slash)_{ba}\gamma_5\bar{H}_a]+h.c.\}
\nonumber\\&& +\{i\frac{h_2}{\Lambda_{\chi}}{\rm
Tr}[T_b^{\mu}(D\!\!\!\!/A_{\mu})_{ba}\gamma_5\bar{H}_a]+h.c.\}+g_\sigma{\rm
Tr}[H \sigma\overline{H}],\label{aa}
\end{eqnarray}
where $H_a=\frac{1+\not v}{2 }[P_{a}^{*\mu}-P_a \gamma_5]$,$
S_{a}=\frac{1+\not v}{2 }[P_{1a}^{'\mu}\gamma_{\mu}\gamma_5
-P_{0a}^{*}]$ and $T_{a}^{\mu}=\frac{1+\not v}{2
}\big\{P^{*\mu\nu}_{2a}
\gamma_{\nu}-\sqrt{\frac{3}{2}}P_{1a}^{\nu}\gamma_5 [g_{\nu}^{\mu}
-\frac{1}{3}\gamma_{\nu}(\gamma^{\mu}-v^{\mu})]\big\}$. The axial
vector field $A_{ab}^{\mu}$ is defined as $
A_{ab}^{\mu}=\frac{1}{2}(\xi^{\dag}\partial^{\mu}\xi-\xi\partial^{\mu}\xi^{\dag})_{ab}$
with $\xi=\exp(i\mathcal{M}/f_{\pi})$, $f_\pi=132$ MeV and
$\mathcal{M}$ is the octet pseudoscalar matrix.

We impose the constraint on the scattering amplitudes that initial
states and final states should have the same angular momentum. The
molecular state $|J,J_z\rangle$ composed of the $1^-$ and $1^+$
charm meson pair can be constructed as
\begin{equation}
|J,J_z\rangle=\sum_{\lambda_1,\lambda_2}\langle
1,\lambda_1;1,\lambda_2|J,J_z \rangle|p_1,\epsilon_1;
p_2,\epsilon_2\rangle
\end{equation}
where $\langle1,\lambda_1;1,\lambda_2|J,J_z\rangle$ is the
Clebsch-Gordan coefficient. Combining the equation with the
scattering amplitudes, one gets the matrix element
$i\mathcal{M}(J,Jz)$.

With the Breit approximation, the interaction potential in the
momentum space is related to $i\mathcal{M}(J,Jz)$
\begin{equation}
V(q)=-\frac{1}{\sqrt{\prod_{i}2m_i \prod_{f}2m_f
}}\mathcal{M}(J,Jz)
\end{equation}
where $m_i$ and $m_f$ denote the masses of the initial and final
states respectively. Then we average the potential in the momentum
space. Finally we make Fourier transformation to derive the
potential in the coordinate space.

\section{Is $Z^+(4430)$ a loosely $D_1'-D^*$ or $D_1-D^*$ molecular state? }

If $Z^+(4430)$ is a $D_1'D^*$ or $D_1D^*$ molecular state, the
flavor wave function of $Z^+(4430)$ is
\begin{eqnarray*}
|Z^{+}\rangle&=&\frac{1}{\sqrt{2}}\Big(|\bar{D}_{1}^{'0}
D^{*+}\rangle+|\bar{D}^{*0}D_{1}^{'+}\rangle\Big),\;\;\;\mathrm{or}\;\;\;|Z^{+}\rangle=\frac{1}{\sqrt{2}}\Big(|\bar{D}_{1}^{'0}
D^{*+}\rangle+|\bar{D}^{*0}D_{1}^{'+}\rangle\Big).
\end{eqnarray*}
For the  flavor wave function of $\tilde{Z}^+$ with opposite
G-party, we only replace the plus sign in the above functions with
a minus sign \cite{xiangliu}.

We only consider the contribution from OPE and obtain the
potentials of $D_1'D^*$ and $D_1D^*$ systems in the coordinate
space, which are listed in Table \ref{aa}.

{\tiny
\begin{center}\begin{table}[htb]\caption{The one pion exchange potential
in the coordinate space with $A'=\bar{D}_{1}'^{0}D^{*+}$,
$B'=D_{1}'^{+}\bar{D}^{*0}$,  $C'=\bar{D}_{1}^{0}D^{*+}$ and
$D'=D_{1}^{+}\bar{D}^{*0}$. Here
$\zeta=\delta(\mathbf{r})-\frac{m_\pi^2}{4\pi r}e^{-m_\pi r}$,
$\eta=\frac{\cos(\mu r)}{r}$,
$\xi=\delta(\mathbf{r})-\frac{m_\pi^2}{4\pi r}e^{-m_\pi r}$,
$\chi=\nabla^2\delta(\mathbf{r})
-\mu^2\delta(\mathbf{r})-\frac{\mu^4}{4\pi}\frac{\cos \mu r}{r}$.
\label{t1}}
\begin{tabular}{c|cc|cc}
\hline
 &\multicolumn{2}{c}{$D_1'-D^*$
system}&\multicolumn{2}{c}{$D_1-D^*$ system} \\\hline
&$A^\prime(B^\prime)\rightarrow A^\prime(B^\prime) $&{$A^\prime
(B^\prime)\rightarrow
B^\prime(A^\prime)$}&$C^\prime(D^\prime)\rightarrow
C^\prime(D^\prime)
$&{$C^\prime (D^\prime)\rightarrow D^\prime(C^\prime)$}\\
\hline\hline $0^-$&$ \frac{gg'}{3 f_\pi^2}\zeta $ &$\frac{h^2
(q^0)^2}{8\pi f_\pi^2}\beta$ &$ -\frac{5gg''}{18 f_\pi^2}\xi
$&$\frac{h'^2}{6 f_\pi^2}\chi$
\\
\hline $1^-$&$\frac{gg'}{6 f_\pi^2}\zeta$&$\frac{h^2 (q^0)^2}{8\pi
f_\pi^2}\beta$&$ -\frac{5gg''}{36 f_\pi^2}\xi $&$-\frac{h'^2}{12
f_\pi^2}\chi$\\ \hline $2^-$&$-\frac{gg'}{6
f_\pi^2}\zeta$&$\frac{h^2 (q^0)^2}{8\pi f_\pi^2}\beta$&$
\frac{5gg''}{36 f_\pi^2}\xi $&$\frac{h'^2}{60f_\pi^2}\chi$\\
\hline
\end{tabular}
\end{table}\end{center}}
Where $g=0.59\pm 0.07\pm0.01$ can be extracted by fitting the
experimental width of $D^*$ \cite{CLEO}. In quark model, Falk and
Luke give an approximate relation $|g'|=|g|/3$ and $|g''|=|g|$
\cite{falk}. With the available experimental information,
Casalbuoni and collaborators extracted $h=-0.56\pm 0.28$ and
$h'=(h_1+h_2)/\Lambda_{\chi}=0.55$ GeV$^{-1}$ \cite{casalbuoni}.
Besides those coupling constants, other parameters include:
$m_{D^*}=2007$ MeV, $m_{D_1'}=2430$ MeV, $m_{D_1}=2420$ MeV,
$m_{B^{*}}=5325$ MeV, $m_{B_{1}'}=5732$ MeV, $f_{\pi}=132$ MeV,
$m_{\pi}=135$ MeV \cite{PDG}; $m_{B_1}=5725$ MeV \cite{PDG-1}.

With the potentials derived above, we use the variational method
to investigate whether there exists a loosely bound state. Our
criteria of the formation of a possible loosely bound molecular
state is (1) the radial wave function extends to 1 fm or beyond
and (2) the minimum energy of the system is negative. Our trial
wave functions include (a) $\psi(r)=(1+\alpha r)e^{-\beta r}$; (b)
$\psi(r)=(1+\alpha r^2) e^{-\beta r^2}$; (c) $\psi(r)=
r^2(1+\alpha r) e^{-\beta r}$.

Unfortunately we does not find a solution to satisfy the above
criteria for the system of $D^{'}_1-D^*$ or $D_1-D^*$ in all
$J^P=0^-, 1^-, 2^-$ channels with the realistic coupling constants
$g=0.59$, $g'=g/3$ and $g''=g$ deduced from the width of $D^\ast,
D_1$ and $D_1^\prime$. Such a solution also does not exist if we
switch the sign of $gg'$ or enlarge the absolute value of $gg'$ by
a factor 3. The same conclusion holds for the system of $B_1'B^*$,
$B_1B^*$ and $\widetilde{Z}^+$ with negative $G$-parity.

It's interesting to note that the one pion exchange potential
alone does not bind the deuteron in nuclear physics either.  In
fact, the strong attractive force in the intermediate range is
introduced in order to bind the deuteron, which is sometimes
modeled by the sigma meson exchange. One may wonder whether the
similar mechanism plays a role in the case of $Z^+(4430)$ and
$X(3872)$. Further work along this direction is in progress
\cite{liuxiang}. Basing on the above considerations, we reanalyze
$DD^*$ system.

\section{Is $X(3872)$ really a $D-D^*$ molecular state?}

We reanalyze the flavor function of $X(3872)$ and obtain
\begin{eqnarray}\label{Xwave}
|X(3872)\rangle=\frac{1}{\sqrt{2}}\Big[|D^0
\bar{D}^{*0}\rangle-|{D}^{*0}\bar{D}^0 \rangle\Big],
\end{eqnarray}
which is naturally reflect the positive C-parity of $X(3872)$.

With the convention of the $X(3872)$ flavor wave function in Eq.
(\ref{Xwave}), the potential in the study of the molecular picture
finally reads as
\begin{eqnarray}\label{Xpotential}
V(\mathbf{r})=g_\sigma^2 Y_\sigma(\mathbf{r})+\frac{g^2}{6f_\pi^2}
Y_\pi(\mathbf{r})
\end{eqnarray}
with $Y_\sigma(\mathbf{r})=\frac{1}{4\pi r}e^{-m_\sigma r}$ and
$Y_\pi(\mathbf{r})=-\delta(\mathbf{r})-\frac{\mu^2}{4\pi
r}\cos(\mu r)$, where $\mu=\sqrt{q_0^2-m_\pi^2}$. The sign between
one sigma exchange potential (OSEP) and OPEP is determined by the
relative sign of $|D^0 \bar{D}^{*0}\rangle$ and
$|{D}^{*0}\bar{D}^0 \rangle$ in the wave function in Eq.
(\ref{Xwave}).

Due to the existence of the three dimensional $\delta$ function in
the potential, Suzuki argued that $D$ and $\bar{D}^\ast$ could not
be bound as a molecular state \cite{suzuki}. We note that the
potential in Eq. (\ref{Xpotential}) is derived with the implicit
assumption that all the mesons are point-like particles. Such an
assumption is not fully reasonable due to the structure effect in
every interaction vertex. Thus in the following we will introduce
a cutoff to regulate the potential and further study whether it is
possible to find a loosely bound molecular state using the
realistic potential. We adopt two approaches: (1) considering the
form factor (FF) contribution; (2) smearing the potential.
Although these two approaches look different, they are essentially
the same, i.e. imposing a short-distance cutoff to improve the
singularity of the effective potential.

With introducing monopole FF
$F(q)=(\Lambda^2-m^2)/(\Lambda^2-q^2)$ in the potential as an
example, we give the modified potential as
\begin{eqnarray}
Y_\sigma(r)&=&\frac{1}{4\pi r}(e^{-m_\sigma r}-e^{-\Lambda
r})-\frac{\eta'^2}{8\pi\Lambda}e^{-\Lambda r},\;
Y_\pi(r)=-\frac{\mu^2}{4\pi r}[\cos(\mu r)-e^{-\alpha
r}]-\frac{\eta^2\alpha}{8\pi}e^{-\alpha r},\nonumber
\end{eqnarray}
where $\eta=\sqrt{\Lambda^2-m_\pi^2}$,
$\eta'=\sqrt{\Lambda^2-m_\sigma^2}$ and
$\alpha=\sqrt{\Lambda^2-q_0^2}$. Note we use the same $\Lambda$
for $\pi$ and $\sigma$ exchange. We found that the $\sigma$
exchange potential is repulsive.

One gets numerical solutions depicted in Table \ref{FFpi_sig}. We
only use the coupling constant $g_\sigma=0.76$ to illustrate the
results. We chose the solutions with $-5.0 \,{\rm MeV}<E_0< -0.1
\,{\rm MeV}$.
\begin{table}[htb]
\centering
\begin{tabular}{c|cccc}\hline
&$\Lambda$ (GeV)&$E_0$ (MeV) &$r_{\rm rms}$ (fm) & $r_{\rm max}$
(fm)
\\\hline  & 6.0 & -1.3 & 2.8 & 0.1
\\\raisebox{1.3ex}[0pt]{$g=0.59$}& 6.1 & -4.9 & 1.5 & 0.1
\\\hline   & 3.3 & -0.7 & 3.8 & 0.3
\\\raisebox{1.3ex}[0pt]{$g=0.8$} & 3.4 & -3.7 & 1.7 & 0.2
\\\hline  & 2.1 & -0.3 & 5.9 & 0.4
\\\raisebox{1.3ex}[0pt]{$g=1.0$} & 2.2 & -2.4 & 2.2 & 0.3
\\\hline
\end{tabular}
\caption{Solutions for various $g$ and $\Lambda$ in the case of FF
with total potential. Lowest eigenvalues between -5.0 MeV and -0.1
MeV are selected. Here $g_\sigma=0.76$ is used. Here $r_{\rm rms}$
is the root-mean-square radius, and $r_{\rm max}$ is the radius
corresponding to the maximum of the wave function $\chi(r)$.}
\label{FFpi_sig}
\end{table}
\begin{table}[htb]
\centering
\begin{tabular}{c|cccc}\hline
&$\Lambda$ (GeV)&$E_0$ (MeV) &$r_{\rm rms}$ (fm) & $r_{\rm max}$
(fm)
\\\hline  & 2.5 & -0.5 & 2.7 & 0.3
\\\raisebox{1.3ex}[0pt]{$g=0.59$}& 2.6 & -2.5 & 1.2 & 0.2
\\\hline
\end{tabular}
\caption{Solutions for various $g$ and $\Lambda$ in the case of FF
for the $B\bar{B}^\ast$ system with the total potential. The
lowest eigenvalues between -5.0 MeV and -0.1 MeV are selected.
Here $g_\sigma=0.76$ is used.} \label{a}
\end{table}
We found (1) $DD^*$ interaction through one pion and one sigma
exchanges is not attractive enough to form a bound state with
$g=0.59$ and $g_{\sigma}=0.76$ and $\Lambda=1$ GeV; (2) when $g$
becomes larger, the critical point for $\Lambda$ to generate a
$DD^*$ bound state becomes small. The $BB^*$ system also is
investigated. The results are shown in Table \ref{a}.

The results for the case of smearing also confirm the above
observation for the case of FF.

\section{Summary}

In a short summary, our numerical results indicate that it is hard
to explain $Z^+(4430)$ as a $D_1'D^*$ or $D_1D^*$ only considering
OPEP without cutoff. However this conclusion needs to be confirmed
by considering OSEP and adding cutoff in potential, which is in
progress. Then a decisive conclusion about whether $Z^+(4430)$ can
be understood as a $D_1'D^*$ or $D_1D^*$ molecular state can be
made.

$X(3872)$ can not be explained as a $DD^*$ molecular state by
considering one pion and one sigma exchanges, and introducing
cutoff in the potential. We also find that there exists an S-wave
$BB^*$ system with $J^P=1^{++}$, which can be searched in further
experiment.

\vfil

{\bf Acknowledgments.}We enjoy the collaboration with Professor
Shi-Lin Zhu. This project was supported by the National Natural
Science Foundation of China under Grants 10625521, 10675008,
10705001, 10775146, 10721063 and the China Postdoctoral Science
foundation (20060400376, 20070420526). X.L. specially thanks the
support of the \emph{Funda\c{c}\~{a}o para a Ci\^{e}ncia e a
Tecnologia of the Minist\'{e}rio da Ci\^{e}ncia, Tecnologia e
Ensino Superior} of Portugal (SFRH/BPD/34819/2007).

\end{document}